# Feasibility study of thermal losses conversion into Light for High Power LEDs using thermoelectric modules

R. Ben Hassen[2], L. Canale[1], Z. Araoud[2], G. Zissis[1]

[1]LAPLACE, Université de Toulouse, CNRS, INPT, UPS, 118, rte de Narbonne 31062 Toulouse Cedex 9, France.
[2]Department of Energy, ENIM, 5000 Street IBN JAZZAR, Monastir 5035, TUNISIA
Contact: laurent.canale@laplace.univ-tlse.fr

**ABSTRACT**

Light Emitting Diodes emits no IR and no UV and their spectrum is fully in the visible part. But LEDs are not cold and all energy losses are thermal losses. The aim of this paper is to prove the feasibility to reuse the thermal losses to produce light through a thermoelectric module. Papers where Peltier modules are included in LEDs systems are all the time used for cooling [1-6]. At the knowledge of the authors, this the first time that thermal losses are used to increase the global efficiency of a high power LED lighting system by using Peltier modules to produce light.

**INTRODUCTION:**

Light Emitting Diodes (LEDs) are one of the most efficient light source on the market. Even though they are much more efficient than traditional light sources, they transform about 60 to 70% of the electrical energy consumed into heat. The function of a LED is to produce light. Thus, every loss transformed into light must increase the efficiency of the system. To prove this concept, we have chosen a high power LED (Bridgelux W3500). After a full thermal modelling of this Chip-On-Board LED to evaluate the thermal losses and predict the power available through Peltier modules, a complete and simple electronic system is realized to verify the predictions.

**THERMAL MODELLING & COMSOL SIMULATION:**

A high power LED is a component that generates high thermal losses. This part describes a full COMSOL modelling of the thermal exchange between all layers of the system. The full knowledge of the heat transfers from the back side of the LED through a heat-sink to the air is the requisite step before including the Peltier Module.

Thermal effects affect the efficiency of the LED and more important, the lifetime. If the main priority stills the thermal management, thermal effects are unavoidable. This is why one of the most critical parts of the system is thermal interface materials (TIMs) that connect the LED to the heatsink. The choice of this material can be crucial for optimum operation. In order to properly choose the TIMs, the criterion is very practical: the thermal resistance. This diagram presents a thermal network for the LED with a heat sink high-lighting, the exchanges that occur between the different interfaces separating the various elements: conduction and convection

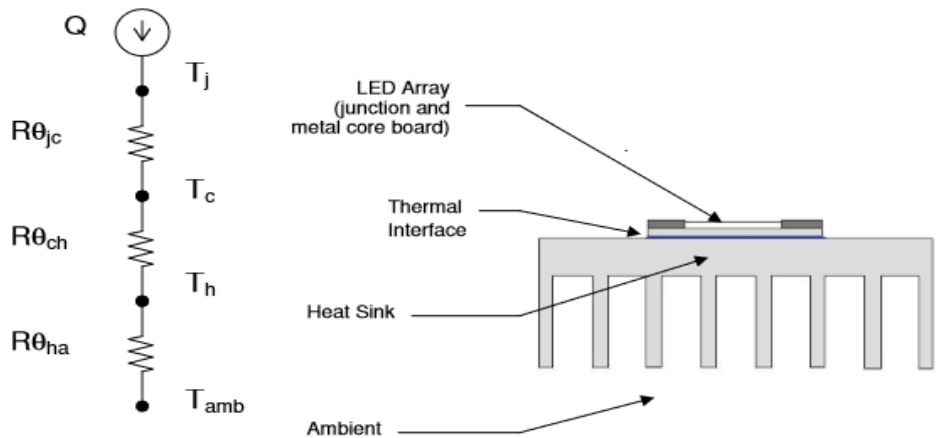

**Figure 1:** Thermal network model of a LED [7]

Where:
Q is heat flowing from hot to cold through the LED
Tj and Tc are the temperature at the junction and the temperature at the case of the LED, respectively
Th is the temperature at the point where the heat sink is attached to the LED
Tamb is the ambient air temperature
$R_{\theta jc}$ is the thermal resistance from junction to case of the LED
$R_{\theta ch}$ is the thermal resistance between the case of the LED and the heat sink
$R_{\theta ha}$ is the thermal resistance of the heat sink
And based on these two equations:

$$R_{\theta,J-amb} = \frac{T_j - T_{amb}}{P_d} \qquad [1]$$
$$R_{\theta,j-amb} = R_{\theta,j-c} - R_{\theta,c-h} - R_{\theta,h-amb} \qquad [2]$$

The bridgelux LED (ref. BXRA-W3500) [8] studied is composed of 8 parallel networks consisting of 8 junctions in series, so the LED module is the equivalent of 64 LEDs junctions. Each LED has a power of about 1 W and the LED is the equivalent of 64 W. These LEDs are concentrated in a small area, so they heat each other, moreover the power per m² to dissipate is more and more important, the temperature in the LED component can be very high, hence the interest of adding a thermoelectric generator TEG that will play the role of a converter using the Peltier effect to convert thermal energy into electrical energy and the rest of heat will be evacuated to ambient air through the heat sink.

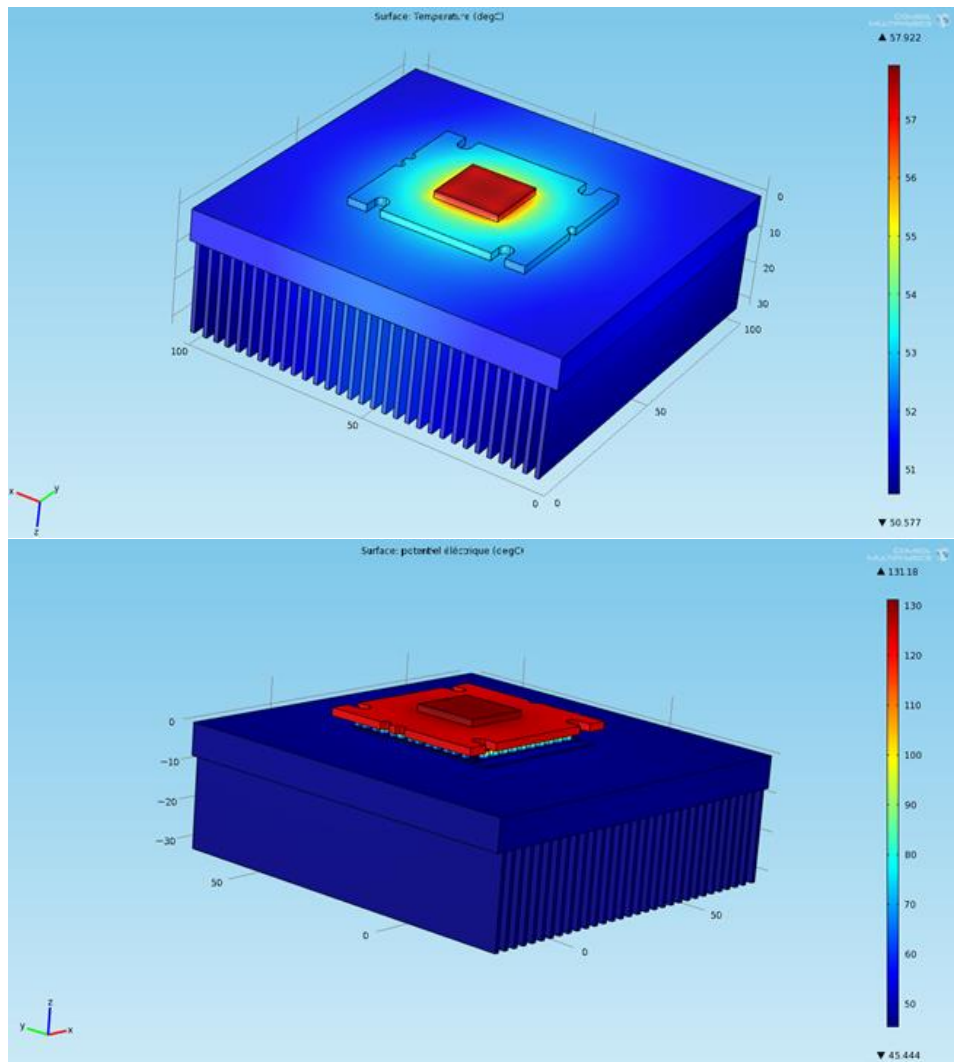

**Figure 2:** COMSOL simulation without (top) and with (bottom) the thermoelectric module

## EXPERIMENTAL SETUP:

- **First test: with a single thermoelectric module**

The first prototype first was a success, we managed to obtain a voltage (although minimal) V = 1V with an intensity I = 300 mA, from the heat dissipated by the LED Bridgelux. The voltage was under the threshold voltage of a red LED (1.6V). Thus, we use another thermoelectric module associated in series with the first one to increase the voltage generated and properly light the LEDs. So, all that remains is to improve this experiment with another thermoelectric module and to produce even more voltage and intensity.

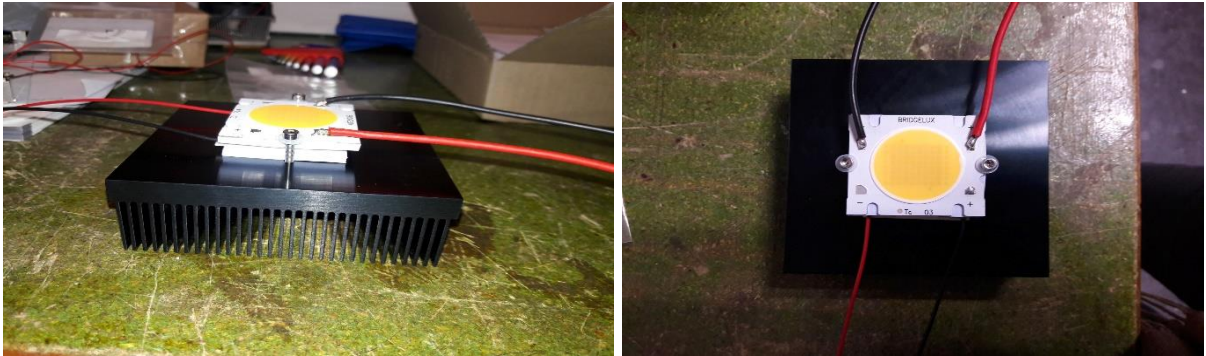

**Figure 3:** The first prototype LED + TEG1+dissipator

- **Second test: with two thermoelectric modules connected in series**

Different configurations are tested and the second experiment illustrated in Figure 3 with two thermoelectric modules allows us to reach a current of I = 332 mA and a voltage V = 1.622 V. This configuration is available to enlighten 25 red LEDs in parallel

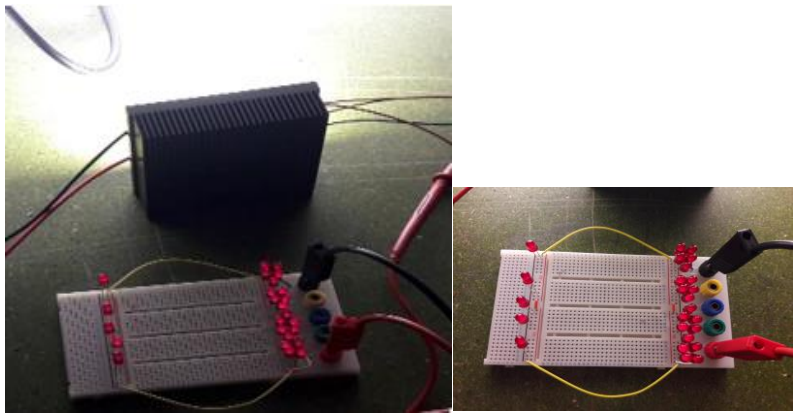

**Figure 4:** Experimental setup with two thermoelectric modules
After few minutes (right side), LEDs shows a strong brightness

- **Third test: With white LEDs**

To allow us to reach a voltage over the threshold of a white LED, we have built a Boost electronic circuit (Fig.5). This Boost electronics will be inserted between the thermoelectric module and the LEDs to be lit.

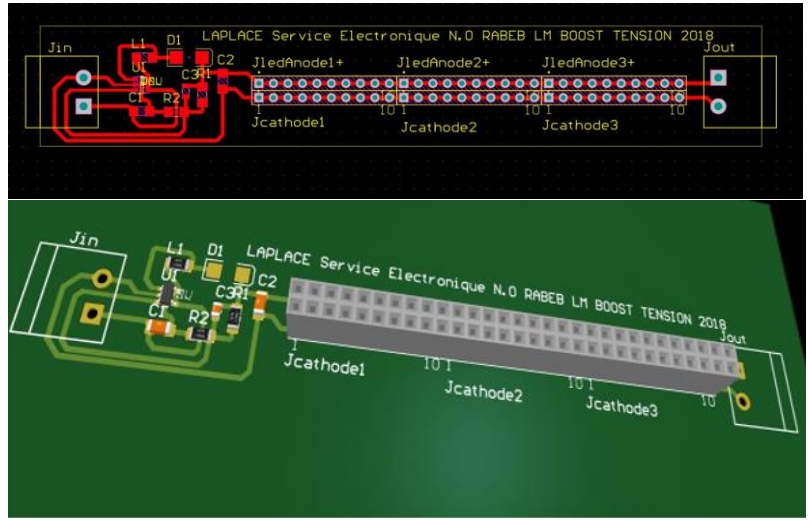

**Figure 4:** Boost Converter DC-DC in 3D (schematic and 3D view)

## CONCLUSION:

After these three experimental tests, a study of feasibility was proved: we have obtained good results and reach to produce more light through thermal conversion. At the knowledge of the author, this paper present the first study using thermoelectric modules to produce light from the heat dissipated by a High Power LED. Nevertheless, even if the effectiveness of such a system remains probably very low, this study is a first. Characterizations in integrating sphere will complete these results to estimate the gain on the luminous efficiency.